\documentclass[%
  preprint,12pt,onecolumn,
  superscriptaddress,
  amsmath,amssymb,
  aps,prx,
  doublespa
]{revtex4-2}

\usepackage{graphicx}      
\usepackage{xcolor}        
\usepackage{mathtools}     
\usepackage{bm}            
\usepackage{bbm}           
\usepackage{multirow}      
\usepackage{enumitem}      
\usepackage{cancel}        
\usepackage[section]{placeins} 
\usepackage{float}         

\usepackage{algorithm}
\usepackage{algpseudocode}

\usepackage{xr-hyper}
\usepackage{hyperref}
\def\circled#1{{\ooalign{\hfil\lower.1ex\hbox{#1}\hfil\crcr\Orb}}}

\begin{document}
\title{Single-Trajectory Bayesian Modeling Reveals Multi-State Diffusion of the MSH Sliding Clamp}

\author{
Seongyu Park\textsuperscript{1,2,*}, 
Inho Yang\textsuperscript{1,2,*}, 
Jinseob Lee\textsuperscript{3}, 
Sinwoo Kim\textsuperscript{1},\\
Juana Martín-López\textsuperscript{4}, 
Richard Fishel\textsuperscript{4}, 
Jong-Bong Lee\textsuperscript{1,3,\dag}, 
Jae-Hyung Jeon\textsuperscript{1,5,\dag}\\[1em]
\textsuperscript{1}Department of Physics, Pohang University of Science and Technology (POSTECH), Pohang 37673, Republic of Korea\\
\textsuperscript{2}Institute for Theoretical Science, POSTECH, Pohang 37673, Republic of Korea\\
\textsuperscript{3}Division of Interdisciplinary Bioscience and Bioengineering, POSTECH, Pohang 37673, Republic of Korea\\
\textsuperscript{4}Department of Cancer Biology and Genetics, The Ohio State University Wexner Medical Center, Columbus, Ohio, USA\\
\textsuperscript{5}Asia Pacific Center for Theoretical Physics (APCTP), Pohang 37673, Republic of Korea\\[1ex]
*These authors contributed equally. \quad
\dag Corresponding authors: jeonjh@gmail.com, jblee@postech.ac.kr 
}\noaffiliation

\begin{abstract}
DNA mismatch repair (MMR) is the essential mechanism for preserving genomic integrity in various living organisms. In this process, MutS homologs (MSH) play crucial roles in identifying mismatched basepairs and recruiting downstream MMR proteins. The MSH protein exhibits distinct functions and diffusion dynamics before and after the recognition of mismatches while traversing along DNA. An ADP-bound MSH, known as the MSH searching clamp, scans DNA sequences via rotational diffusion along the DNA backbone. Upon recognizing a mismatch, the MSH combines with ATP molecules, forming a stable sliding clamp. 
Recent experimental evidence challenges the conventional view that the sliding clamp performs a simple Brownian motion.
In this study, we explore the diffusion dynamics of the ATP-bound MSH sliding clamp through single-particle tracking experiments and a Bayesian diffusion-state analysis method. Our quantitative analysis reveals that the diffusion characteristics defy explanation by a single-state diffusion mechanism. Instead, our in-depth model inference uncovers three distinct diffusion states, each characterized by specific diffusion coefficients: $D_1=1.86 \times 10^{-2}~\mathrm{\mu m^2/s}$, $D_2=1.30 \times 10^{-1}~\mathrm{\mu m^2/s}$, and $D_3=9.64 \times 10^{-1}~\mathrm{\mu m^2/s}$, respectively. 
These three states alternate over time, with cross-state transitions predominantly involving $D_2$-state, and direct transitions between $D_1$- and $D_3$-states being scarce.
We propose that these multi-state dynamics reflect underlying conformational changes in the MSH sliding clamp, highlighting a more intricate diffusion mechanism than previously appreciated. 
\end{abstract}
\maketitle



\section{Introduction}
DNA mismatch repair (MMR) is a crucial post-replication process that guarantees genomic integrity. Erroneous DNA segments, like base-base mispairs (mismatches) and insertion-deletion loops, occur approximately once per million base pairs during DNA replication~\cite{fishel2015mismatch}. Failure to correct these erroneous segments can result in severe diseases such as Lynch syndrome~\cite{wheeler2000dna, muller2002mismatch}. Upon the mismatches in DNA, a group of proteins called mismatch repair proteins is incorporated to perform a series of MMR processes. The commonly accepted scenario regarding the MMR process unfolds as follows: ADP-bound MutS homologs (MSH) initiate the MMR process by recognizing a lesion site over one-dimensional rotation-coupled diffusion along the DNA backbone. Upon binding to the mismatched nucleotide, the MSH prompts the exchange of ADP, initially bound to the ATPase domain of MSH, with ATP. This nucleotide exchange induces a conformational change in MSH, transitioning it into a stable, hydrolysis-independent sliding clamp (as depicted in Fig.~\ref{fig1}(c))~\cite{cho2012, jeong2011muts, fishel2015mismatch, gradia1999hmsh2}. 
The ATP-bound MSH sliding clamp, while freely diffusing around the mismatch site without re-engaging it, triggers the loading of MutL homologs (MLH) onto the DNA~\cite{london2021linker}. This MLH loading on the DNA acts as a mediator for subsequent MMR processes, enabling interactions with other MMR proteins~\cite{liu2018stochastic}. Yet, numerous aspects concerning the mechanisms of MSH remain enigmatic and controversial. These contentious topics encompass the conformation of the MSH sliding clamp, the precise timing of ATP involvement in post-mismatch recognition steps, and the necessary number of ATPs essential for the formation of an MSH sliding clamp~\cite{putnam2020muts, wang2016long}.

Contrary to the traditional notion that the MSH sliding clamp maintains a stable conformation, a recent FRET experiment for Thermus aquaticus (Taq) MutS observed multiple, seemingly ATP-dependent FRET signals after the formation of the sliding clamp~\cite{hao2020recurrent}. This observation suggests that the MSH sliding clamp may undergo conformational transitions related to ATP association, dissociation, or hydrolysis~\cite{putnam2020muts}. However, other studies have not observed such multiplicity in states, including mismatch rebinding events by the MSH sliding clamp~\cite{cho2012, london2021linker, liu2016cascading}. Therefore, a deeper quantitative understanding of the conformations of the MSH sliding clamp, such as the number of existing states and the transition dynamics among these states, remains to be established. 

In this study, we address this challenge by analyzing the diffusion dynamics of the MSH sliding clamp on DNA. We conducted single-particle tracking (SPT) experiments using the DNA skybridge platform~\cite{kim2019dna}. Our extensive statistical analysis of the 1D diffusion trajectories revealed that the diffusion dynamics of MSH sliding clamps do not follow simple Brownian motion; instead, they exhibit time-varying properties. Recent studies on the 1D diffusion of DNA-binding proteins have shown that several DNA-binding proteins diffuse with temporally varying mobility due to conformational changes~\cite{beckwitt2020single, cheon2019single, yamamoto2021universal, park2021mini}. Consistent with these findings, we hypothised that the observed time-varying behavior is linked to the conformational variability of MSH sliding clamps. 

To quantitatively characterize the time-varying diffusion state, we devised a comprehensive workflow for diffusion-state analysis. Our diffusion-state analysis machine involves determining the number of diffusion states, their respective diffusion coefficients, transition probabilities between states, and identifying time-dependent diffusion states along a trajectory. Our framework is primarily grounded in Bayesian inference~\cite{thapa2018, park2021bayesian}, offering interpretable results compared to deep-learning methods. 

Applying our diffusion-state analysis method to the single-particle trajectory data of MSH sliding clamps, we unveil that the diffusion of an ATP-bound MSH sliding clamp exhibits three dynamic states with distinct diffusion coefficients of $D^{(1)}\sim 1.86 \times 10^{-2} \mathrm{\mu m^2/s}$, $D^{(2)}\sim 1.30 \times 10^{-1} \mathrm{\mu m^2/s}$, and $D^{(3)}\sim 9.64 \times 10^{-1} \mathrm{\mu m^2/s}$. The diffusion states frequently transition between different states on a timescale of a few seconds. In addition, it appears that the diffusion state characterized by the diffusion coefficient plays a mediating role in transitions between the other two states. These findings challenge the long-held assumption of a single diffusion state~\cite{fishel2015mismatch, cho2012, jeong2011muts} and suggest that frequent conformational changes occur after the initial formation of the MSH sliding clamp. Based on these findings, we propose a diffusion model for the MSH sliding clamp in the absence of MLH-PMS, a downstream effector protein, which reflects a functional state prior to activation of the DNA mismatch repair pathway.

\section{\label{sec:prelim} Heterogeneous diffusion of MSH proteins revealed by single-molecule experiments}

\begin{figure*}[!ht]
\centering
\includegraphics[width=0.85\textwidth]{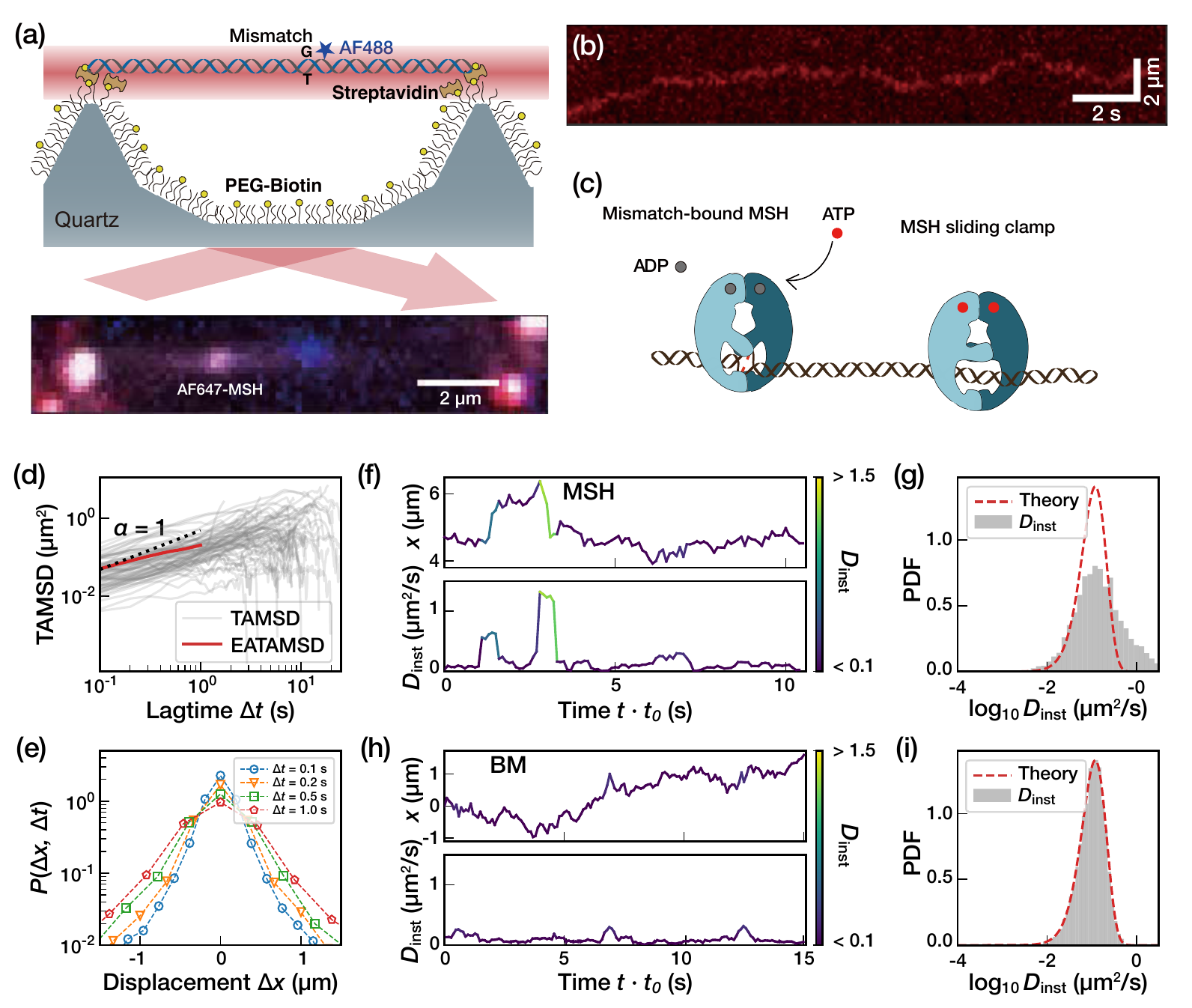} 
\caption{\scriptsize\textbf{One-dimensional diffusion of an MutS homologs (MSH) protein along DNA.} (a) A schematic of the DNA skybridge platform and a snapshot of real-time sm-TIRF imaging showing the movement of an Alexa647 (AF647)-tagged MSH protein along modified $\lambda$-phage DNA molecule containing a G/T mismatch and an AF488 fluorophore positioned 9 nucleotides away from the mismatch. (b) A representative kymograph showing the trajectory of an MSH protein as it moves along the DNA molecule. (c) Schematic illustration of the widely acknowledged conformational change in an MSH protein upon binding to a mismatch site. Following its initial binding to the mismatch site, where an MSH search clamp resides for approximately $5~\mathrm{s}$~\cite{honda2014}, the MSH protein associates with an ATP, leading to a conformational shift that transforms it into a stable sliding clamp.
(d) Time-averaged (TA) MSD curves from individual MSH trajectories (gray) and their ensemble average (EATAMSD) (red). 
TA MSD is defined from a single trajectory via 
$
\overline{\delta^2 (\Delta t)} = \frac{1}{T-\Delta t/t_0}\sum_{t=1}^{T-\Delta t/t_0} (x_{t+\Delta t/t_0} - x_{t})^2
$
where $\Delta t$ and $t_0$ are time lag (s) and time resolution (0.1~$\mathrm{s}$), respectively.
The black dotted line represents the scaling of Fickian diffusion ($\alpha=1$). 
(e) The van-Hove self-correlation function obtained from all MSH trajectories for lag times $\Delta t = 0.1$, $0.2$, $0.5$, and $1~\mathrm{s}$. 
(f) A trajectory $x(t)$ of the MSH clamp (Top) and the corresponding instantaneous diffusion coefficient $D_\mathrm{inst}(t)$ (Bottom). The color code indicates the value of $D_\mathrm{inst}$.
(g) Normalized histogram of the measured $D_\mathrm{inst}$ from MSH trajectories (gray) compared with the theory (red) of Fickian diffusion [\eqref{eq:log_chi2}]  with $D = 0.12~\mathrm{\mu m^2/s}$.
(h) A trajectory $x(t)$ of a computer-generated Brownian motion (BM) (Top) and the corrresponding instantaneous diffusion coefficient $D_\mathrm{inst}(t)$ (Bottom).
(i) Normalized histogram of the measured $D_\mathrm{inst}$ from the simulated BM trajectories (gray) and \eqref{eq:log_chi2} for $D = 0.12~\mathrm{\mu m^2/s}$ (red).
}
\label{fig1}
\end{figure*}

We have performed single-particle tracking experiments for one-dimensional diffusion of the human MSH2-MSH6 dimer (hereafter referred to as MSH proteins) using the sm-TIRF technique~\cite{liu2016cascading} combined with a DNA skybridge surface interference-free light-sheet imaging [Fig.~\ref{fig1}(a)]~\cite{kim2019dna}. The experimental details are described in Materials and Methods. We have imaged the Alexa647-tagged MSH proteins moving along a $\lambda$-phage DNA that contains an artificially made single mismatch site~[Fig.~\ref{fig1}(a) and (b)]. Previous single-molecule studies reported that an ATP-free MSH searching clamp dwells on a bare DNA for only a few seconds, while an ATP-bound MSH sliding clamp, forming a stable clamp, remains bound to the DNA for approximately 
$\sim 190~\mathrm{s}$~\cite{honda2014}. The comparable durations of experimentally observed trajectories typically indicate successful formation of the MSH sliding clamp.

In our experiment [Fig.~\ref{fig1}(a)--(c)], the one-dimensional diffusion dynamics of an MSH sliding clamp is recorded by time series of position $x_t$ (with time index $t$): $X_i = \{ x_1, x_2, \dots\}_i$, where $i$ denotes the index of a trajectory. The time resolution $t_0$ between successive data points is 0.1~s, and the trajectory length varies from 11 to 201. In total, $N_\mathrm{trj}=62$ trajectories are analyzed.

With our SPT data and typical trajectory analysis methods, we carefully examine the diffusion properties of MSH sliding clamps (i.e., the ATP-bound MSH protein). In a nutshell, our analyses below demonstrate shreds of evidence that an MSH sliding clamp has multiple diffusion states, and its diffusion along the DNA is temporally heterogeneous. 

\subsection{MSH proteins exhibit Fickian yet non-Gaussian diffusion}

We examine the diffusion of individual MSH sliding clamps by measuring the time-averaged (TA) mean squared displacement (MSD) from single trajectories. In Fig.~\ref{fig1}(d), we plot TA MSD curves from 62 single trajectories (gray solid lines) together with their ensemble average curve EATAMSD (red solid line). The results reveal heterogeneous diffusion of MSH sliding clamps in the sense that among 62 trajectories only a few TA MSD curves align with the average curve, and the amplitudes of the TA MSDs exhibit significant scattering around this average.
While EATAMSD shows a mild subdiffusive behavior (the anomalous exponent $\alpha\approx 0.9$), the overall diffusion of MSH sliding clamps is nearly Fickian. Our analysis is consistent with the previous studies reporting that the MSD of the MSH sliding clamps linearly increases with time~\cite{cho2012}.

We then estimate the van-Hove self-correlation function $P(\Delta x, \Delta t)$---the probability density that the displacement of an MSH clamp for a given time lag is $\Delta x$ within the bin size. For typical diffusing particles following Einstein's diffusion theory, the van-Hove self-correlation function displays Gaussian statistics.
In Fig.~\ref{fig1}(e) we plot $P(\Delta x, \Delta t)$ estimated from all trajectories $\{X_i \}$ for several values of $\Delta t$. 
Notably, the MSH sliding clamp deviates from the Gaussian diffusion. Instead, their displacements exhibit an exponentially decaying distribution 
\begin{eqnarray}\label{eq:laplace}
P(\Delta x, \Delta t) \propto \exp \left ( -\mathcal{C}\frac{|\Delta x|}{\Delta t} \right ),
\end{eqnarray}
particularly, at large displacements.

In recent years, a number of experimental and computational studies have unveiled soft matter and biological systems displaying such exponentially decaying van-Hove self-correlation functions. According to these studies, such non-Gaussian diffusion can be attributed to two distinct physical origins. (1) The non-Gaussian diffusion may emerge when a system has multiple diffusion states and its dynamic state (quantified by diffusion coefficient) changes with time. This is called \textit{temporal heterogeneity}. Numerous biological systems exhibit such dynamic features, see the examples ~\cite{metzler2017gaussianity, sposini2018random, wang2012brownian, sabri2020elucidating, chechkin2017brownian, chubynsky2014diffusing, lanoiselee2018diffusion, jeon2016protein, sposini2018random, cherstvy2019non, thapa2018, lanoiselee2018model}. The temporally heterogeneous diffusion coefficient may originate from the spatiotemporal heterogeneity of the surrounding environment or conformational variability of tracer particles~\cite{uneyama2015fluctuation, yamamoto2021universal, miyaguchi2017elucidating, kamagata2018high}. (2) The non-Gaussian PDF originates from particle-to-particle heterogeneous diffusion coefficients while individual particles exhibit an ordinary Gaussian diffusion with a time-independent diffusion coefficient~\cite{sposini2018random}. This is called particle-to-particle heterogeneity, which may occur when individual particles reside in distinct environments or possess varying particle sizes~\cite{thapa2019transient}.
To discern between these two potential scenarios, we conduct additional analyses utilizing deep learning methods (Fig.~\ref{SM-fig_a1} and Supplementary Section~\ref{SM-sec:appendixdeeplearning}).  
The analysis suggests that the annealed transit time model (ATTM) model is the most probable model explaining the MSH trajectory data. The ATTM model describes a temporally heterogeneous diffusion process, where the diffusion coefficient of a Brownian particle changes with time in a step-like manner~\cite{massignan2014nonergodic}.  
Consequently, the deep-learning analysis strongly supports the first scenario of temporally heterogeneous diffusion for the sliding motion of the ATP-bound MSD clamp.

\subsection{Diffusion of MSH proteins is temporally heterogeneous}
To directly visualize the fluctuation in the diffusion coefficient, we measure the instantaneous diffusion coefficient, defined as
$D_\mathrm{inst}(t) = \frac{1}{5}\sum_{t'=t-2}^{t+2} \frac{\Delta x_{t'}^2}{2t_0}$.
Here, $\Delta x_t = x_{t+1} - x_t$ and $t_0$ is the time resolution. Figure~\ref{fig1}(f) presents a representative example of an MSH trajectory and the corresponding $D_\mathrm{inst}(t)$. 
For comparison, Fig.~\ref{fig1}(h) displays the time traces of $x(t)$ and $D_\mathrm{inst}(t)$ for a computer-generated Brownian motion (BM) with a diffusion coefficient $D=0.12~\mathrm{\mu m^2/s}$.

While $D_\mathrm{inst}(t)$ for a BM trajectory smoothly fluctuates around the given value of $D$, the $D_\mathrm{inst}(t)$ for the MSH clamp exhibits significant temporal fluctuations. The difference becomes more apparent when examining the probability density functions (PDFs) of $\log_{10} D_\mathrm{inst}$.

In the case of an ordinary Gaussian diffusion process [Fig.~\ref{fig1}(h)], the instantaneous diffusion coefficient is governed by the theoretical PDF 
\begin{eqnarray}\label{eq:log_chi2}
P(X) = \ln 10 \cdot \log \chi_{\nu=5}^2 \left[ X \ln 10 - \ln \frac{D}{\nu \cdot(\mathrm{\mu m^2/s}) } \right]
\end{eqnarray}
where the argument is $X = \log_{10} \left [ D_\mathrm{inst} \cdot (\mathrm{\mu m^2/s})^{-1} \right ]$ and $\log \chi_{\nu}^2$ represents the log chi-squared distribution with $\nu$ degrees of freedom~\cite{lee2012bayesian}
\begin{eqnarray}
\log \chi_{\nu}^2(Y) = \frac{1}{2^{\nu/2} \Gamma(\nu / 2)} \exp \left [ \frac{1}{2} \nu Y - \frac{1}{2} \exp(Y)\right ].
\end{eqnarray}
As shown, the BM trajectory agrees perfectly with the expected theoretical curve \eqref{eq:log_chi2} [Fig.~\ref{fig1}(i)]. Contrary to this, the estimated PDF for the MSH sliding clamp exhibits a much broader profile than the theoretical log chi-squared distribution [Fig.~\ref{fig1}(g)]. This broadening of $P(\log_{10} D_\mathrm{inst})$ further supports the idea that the diffusion of the MSH sliding clamp does not adhere to a single diffusion mechanism. Moreover, note that identifying diffusion heterogeneity, such as determining how many diffusion states exist or when diffusion-state transitions occur, based solely on the information provided by $D_\mathrm{inst}(t)$ is not a straightforward task. Even for a heterogeneous diffusion process with multiple states, the estimated $P(\log_{10} D_\mathrm{inst})$ turns out to be unimodal and lacks clear clusters [Fig.~\ref{fig1}(g)], making it challenging to establish a threshold value (or the decision boundary) of $D_\mathrm{inst}$ for discerning distinct diffusion states.

Therefore, our above analyses strongly suggest that the MSH sliding clamp exhibits temporally heterogeneous diffusion with multiple diffusion states. This phenomenon aligns with previously reported examples of temporal heterogeneous diffusion, such as particles navigating through a polymer network or a crowded fluid~\cite{chechkin2017brownian, chubynsky2014diffusing, lanoiselee2018diffusion, luo2019quenched, miyaguchi2016langevin, francesco2022, jeon2016protein}. The temporal heterogeneous diffusion is also found in entities that undergo transitions between multiple diffusion states, each associated with distinct conformations~\cite{uneyama2015fluctuation, yamamoto2021universal, miyaguchi2017elucidating, kamagata2018high}. It is plausible that the temporal heterogeneity of the MSH clamp, which does not diffuse within a complex or crowded medium, stems from its occupancy of multiple diffusion states, each characterized by distinct diffusion coefficients.
To quantify such temporally heterogeneous diffusion processes, in the following section, we develop a theory that models heterogeneous diffusion based on a hidden Markov model.

\section{Single-particle Bayesian modeling of heterogeneous MSH diffusion dynamics}

\begin{figure*}[ht]
\centering
\includegraphics[width=\textwidth]{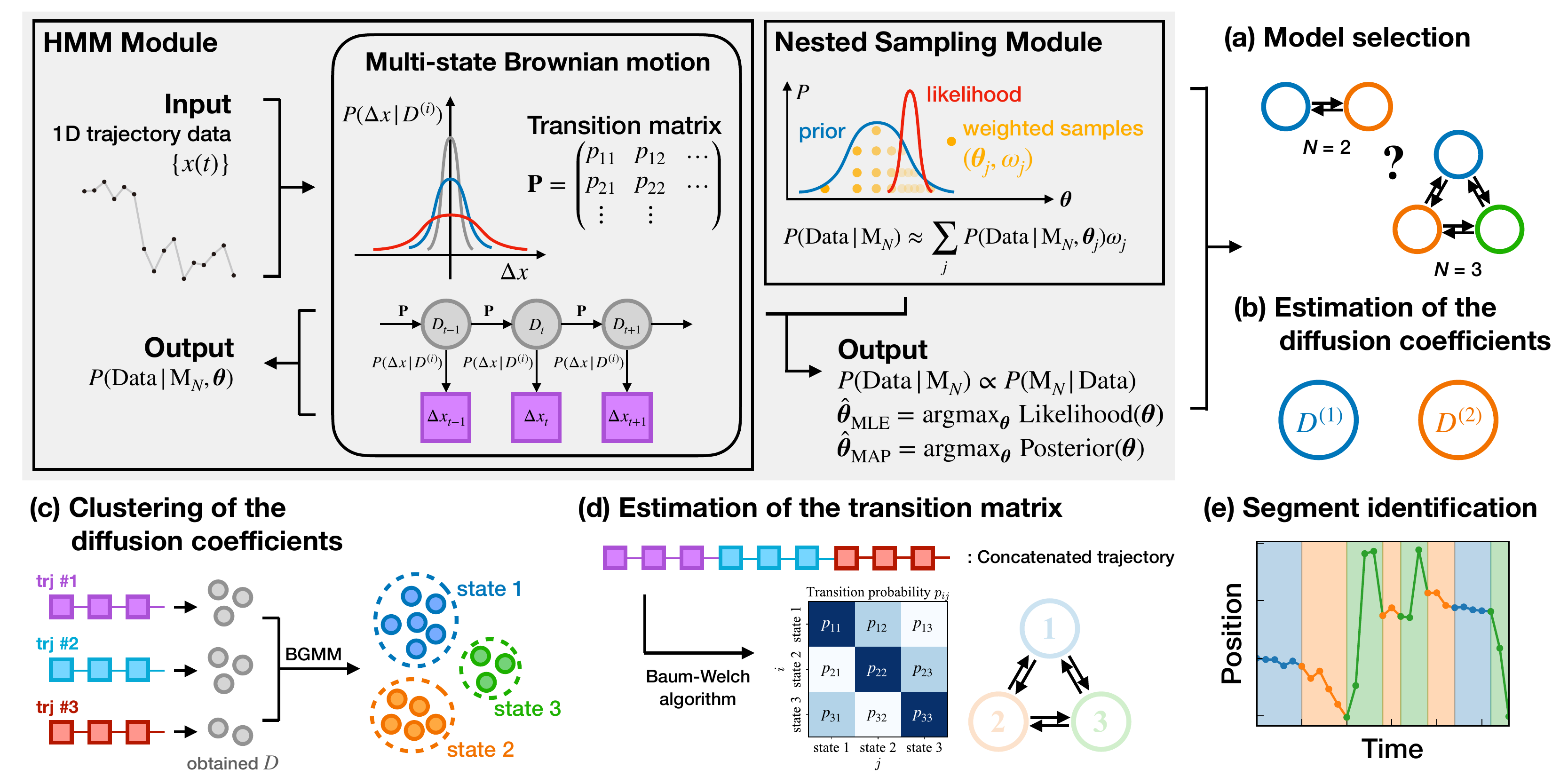} 
\caption{\scriptsize
\textbf{An overview of our diffusion-state analysis method.} 
Single-particle trajectories are analyzed using Bayesian nested sampling integrated with a hidden Markov model of multi-state Brownian motion. The outputs from these modules are subsequently analyzed as follows.
(a) Model selection: Based on the multi-state Brownian motion (BM) model $\mathrm{M}_N$, we infer the optimal model that best fits a given SPT dataset using Bayesian model selection and the Akaike information criterion (AIC) method. The optimal model is trajectory-specific. 
(b) Estimation of diffusion coefficients: Once the optimal BM model $\mathrm{M}_N$ is determined for a given trajectory, we infer the diffusion coefficients of the model, $\{ {D^{(1)},~D^{(2)}, \cdots, D^{(N)}}\}$, using the maximum likelihood estimator (MLE) and maximum a posteriori (MAP) techniques.
(c) Clustering of the obtaine diffusion coefficients: Next, we cluster these diffusion coefficients from all trajectories to determine the total number of distinct diffusion states within our observed data and their associated diffusion coefficients.
(d) Estimation of the transition matrix: From the entire trajectory dataset, we infer the transition probabilities $p_{ij}$ between all combinations of states $i$ and $j$ using MLE ($i,j=1,~2,,\cdots,N$). Additionally, the stationary distribution $\boldsymbol{\pi}$ of the diffusion states is calculated based on the estimated transition matrix. 
(e) Segment identification: Based on the estimated model parameters, we assign the diffusion state $D_t$ of the system as a function of time $t$. The $D_t$ is chosen from the set ${D^{(1)}, D^{(2)}, \dots}$ that maximizes the conditional probability $P(D_t | \Delta x_{1:T}, \hat{\boldsymbol{\theta}})$ at every time step $t$.
}
\label{fig2}
\end{figure*}

To develop a comprehensive tool for identifying and quantifying temporally heterogeneous diffusion processes from SPT trajectory data, we conceive a multi-state diffusion process mimicking the observed sliding diffusion of the ATP-bound MSH clamp. With this diffusion model, we set up probability theories for inferring the number of hidden diffusion states, their diffusion coefficients, and the transition matrix from the data. The workflow of our work is depicted in Fig.~\ref{fig2}.
We validate our methodology by applying it to computer-generated trajectory data of our heterogeneous diffusion model.

\subsection{MSH diffusion is modeled by multi-state Brownian motion}\label{sec:3A}
As a minimal stochastic model, we introduce a Markov chain model called Multi-state Brownian Motion. This is a heterogeneous Brownian particle that has $N$ distinct diffusion states dictated by the diffusion coefficient
\begin{equation}
\mathbf{D} = ( D^{(1)}, D^{(2)}, \dots, D^{(N)} ),
\end{equation}
where the transition between distinct diffusion states follows a Poissonian switching. 
The transition from state $i$ ($D^{(i)}$) to state $j$ ($D^{(j)}$) is dictated by the transition matrix $\mathbf{P}$ whose element  $[\mathbf{P}]_{ij}=p_{ij}$
describes the corresponding transition probability over the unit timestep. In our multi-state BM process, the initial probability of the diffusion state $\boldsymbol{\pi} = (\pi_1, \dots, \pi_N)^\intercal$ is set to be the stationary distribution of the Markov chain, which satisfies the eigenvalue equation $\boldsymbol \pi = \mathbf{P}^\intercal \boldsymbol \pi$. 

The specification of an $N$-state BM model—referred to as $\mathrm{M}_N$—involves a total of $N^2$ independent model parameters. The model parameters are represented by 
\begin{equation} \label{eq:theta}
\boldsymbol{\theta} =\{ D^{(i)}, p_{ij}\}_{i, j},
\end{equation} 
where $i$ runs from 1 to $N$ and $j$ from 1 to $N-1$.
We utilize the multi-state BM model to determine the number of diffusion states of an MSH sliding clamp, estimate the diffusion coefficients for each state, and establish the transition probabilities among these distinct states. 

\subsection{The number of diffusion states inferred using Bayesian model selection and information criteria}\label{sec:3B}

We employ multiple methods for statistical model selection, including those based on Bayesian model evidence and information criteria. 
We briefly explain our model selection and parameter estimation methods. 
A comprehensive technical description of these methods is available in Supplementary Sections~\ref{SM-sec:appendixprior}--\ref{SM-sec:embedding}. 

In the context of Bayesian statistics, the selection of a model is accomplished by comparing the posterior probabilities of various models, denoted as $P(\mathrm{M}_N | \mathrm{Data})$, to identify the most suitable model that maximizes this conditional probability for given data (i.e., single trajectory $\Delta x_{1:T}$). Bayes' theorem provides a practical approach to compute this conditional probability:
$P(\mathrm{M}_N | \mathrm{Data}) = P(\mathrm{Data} | \mathrm{M}_N) P(\mathrm{M}_N)/P(\mathrm{Data})$,
where $P(\mathrm{M}_N)$ is a prior probability (or initial belief) of a model $\mathrm{M}_N$, and $P(\mathrm{Data} | \mathrm{M}_N)$ is a marginal likelihood (or model evidence) of model $\mathrm{M}_N$. In our work, we set the model prior as a flat prior, i.e., $P(\mathrm{M}_I) = P(\mathrm{M}_J)$ for $I \neq J$. The marginal likelihood is obtained via marginalization
\begin{eqnarray}\label{eq:marginalization}
P(\mathrm{Data} | \mathrm{M}_N) = \int_\Theta P(\mathrm{Data}|\mathrm{M}_N, \boldsymbol{\theta}) P(\boldsymbol{\theta}|\mathrm{M}_N)~\mathrm{d} \boldsymbol{\theta}.
\end{eqnarray}
Here, $P(\mathrm{Data}|\mathrm{M}_N, \boldsymbol{\theta}) \equiv \mathcal L(\boldsymbol{\theta} |\Delta x_{1:T}, \mathrm{M}_N)$ is the likelihood function, and $P(\boldsymbol{\theta} | \mathrm{M}_N)$ is a prior distribution of the parameters $\boldsymbol{\theta}$.
The prior distribution is written as $P(\boldsymbol{\theta}|\mathrm{M}_N) = P(\boldsymbol{D}|\mathrm{M}_N) \cdot P(\boldsymbol{P}|\mathrm{M}_N)$, where the mathematical expressions for $P(\boldsymbol{D}|\mathrm{M}_N)$ and $P(\boldsymbol{P}|\mathrm{M}_N)$ are explained in Supplementary Section~\ref{SM-sec:appendixprior}.
The marginalization process in \eqref{eq:marginalization} is technically a difficult step in Bayesian inference. A new method is developed via the nested sampling algorithm, as described in Supplementary Sections~\ref{SM-sec:appendixNS} and \ref{SM-sec:appendixMetropolisGibbs}. Using the estimated Bayesian model evidence, the Bayesian model comparison is finally conducted with $P(\mathrm{M}_N \mathrm{~is~true})=P(\mathrm{Data} | \mathrm{M}_N)/\sum_{\mathrm{M}_I} P(\mathrm{Data} | \mathrm{M}_I)$ where the summations are performed on all candidate models $\mathrm{M}_I$.


To complement and cross-validate our Bayesian model selection results, we additionally perform a model selection analysis using the Akaike Information Criterion (AIC)~\cite{das2009hidden, mckinney2006analysis}. 
In this method, the model with the smallest information criterion value is identified as the best-fit model. For an $N$-state BM model $\mathrm{M}_N$, its AIC is defined as $\mathrm{AIC}_N = 2K_N - 2 \log \mathcal L ( \hat {\boldsymbol{\theta}} _\mathrm{MLE} | \Delta x_{1:T}, \mathrm{M}_N)$~\cite{akaike1974new, schwarz1978estimating},
where $K_N$ denotes the number of (independent) model parameters, $T$ is the sample size, and $\mathcal L ( \hat{\boldsymbol{\theta}} _\mathrm{MLE} |\Delta x_{1:T}, \mathrm{M}_N )$ is the maximum likelihood with the maximum likelihood estimator (MLE), $\hat{\boldsymbol{\theta}}_\mathrm{MLE} = \mathrm{argmax}_{\boldsymbol{\theta}} \mathcal{L} (\boldsymbol{\theta} | \Delta x_{1:T}, \mathrm{M}_N)$.
The maximum likelihood estimator $\hat{\boldsymbol{\theta}}_\mathrm{MLE}$ is obtained through nested sampling (Supplementary Section~\ref{SM-sec:appendixNS}). 

We have tested the performance of our model selection method using simulated trajectories (Supplementary Section~\ref{SM-sec:appendix_modelseleciton}). Our results show that the Bayesian model selection generally outperforms the AIC, while AIC is more effective when the differences between the diffusion coefficients are slight.

\subsection{The estimation of model parameters}

After determining the statistical model $\mathrm{M}_N$ to given SPT data, the next step is to estimate the corresponding model parameters $\boldsymbol{\theta}$, \eqref{eq:theta}, via two distinct estimators: maximum likelihood estimator (MLE) and maximum a posteriori (MAP). The MLE method looks for the parameter set that leads to the maximum likelihood value. For numerical implementation, the nest sampling process is employed in our work. The MAP estimator numerically finds the optimal parameter $\hat{\boldsymbol{\theta}}_\mathrm{MAP}$ that maximizes the posterior distribution
$P(\boldsymbol{\theta}|\mathrm{M}_N, \mathrm{Data}) = P(\mathrm{Data}|\mathrm{M}_N, \boldsymbol{\theta})P(\boldsymbol{\theta}|\mathrm{M}_N)/P(\mathrm{Data}|\mathrm{M}_N)$
using the sampled parameters from the nested sampling.  
For both estimators, the estimated diffusion coefficients are constrained within the range of $D\in[0.001, 2]~(\mathrm{\mu m^2/s})$. 
This constraint is based on the fact that the majority of reported DNA-binding proteins typically exhibit diffusion coefficients within this range~\cite{tafvizi2008, murata2015, murata2017, kamagata2018high, gorman2007, gorman2010visualizing, vestergaard2018, cho2012, jeong2011muts, biebricher2009, cheon2019single, lin2014, lin2016, liu2016cascading}. Moreover, our further investigation reveals that a broader prior distribution results in a decreased performance of model selection (not shown).

\subsection{Clustering of the estimated diffusion coefficients}\label{sec:3d}

The above analysis is based on a single trajectory, so the statistically inferred model and its parameter values may differ from trajectory to trajectory. There is a substantial chance that all distinct diffusion states may not emerge in a single trajectory limited by a finite observation time, potentially leading to underestimation of diffusion states~\cite{chen2016analyzing}. Moreover, if the diffusion coefficients of two distinct dynamic states are similar, accurate differentiation between the two is nontrivial.  We tackle these issues by additionally performing a cluster analysis of the estimated model parameters obtained from a set of individual trajectories.

\begin{figure}
\centering
\includegraphics[width=0.5\textwidth]{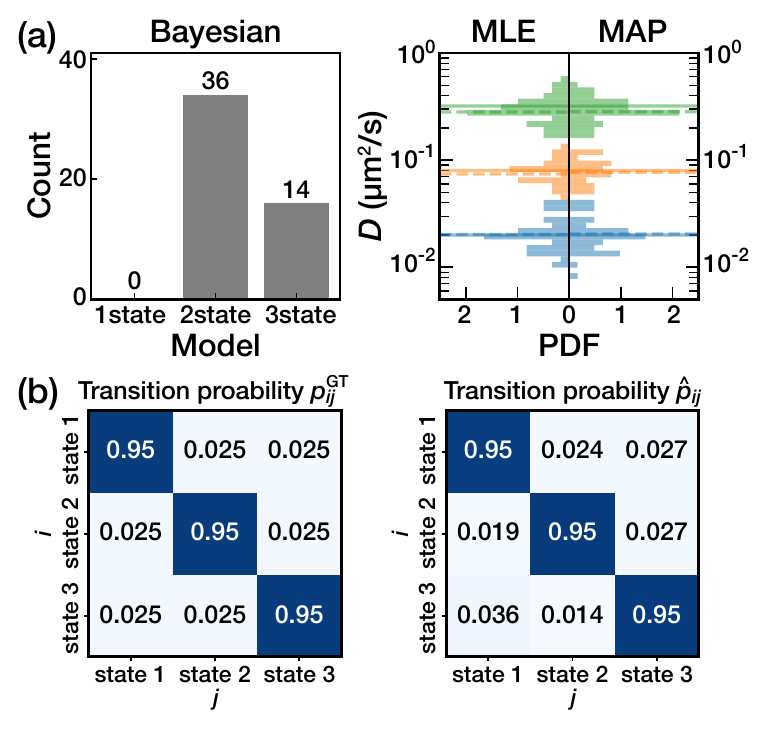} 
\caption{
\textbf{Numerical test of our model inference and parameter estimation methods}. 
The trajectory data used in this analysis are a computer-generated 3-state BM process with the diffusion coefficients $\mathbf{D} = \{0.02, ~0.08, ~0.32\}~\mathrm{\mu m^2/s}$. (a) Model inference is conducted using Bayesian model evidence. The left panel displays the histogram of the inferred diffusion model, and the right panel shows the clustering of the diffusion coefficients.
In the right panel, the bars represent the normalized histograms of the estimates from MLE and MAP, with solid lines marking the ground-truth values and dashed lines denoting the modes of the three Gaussian components identified by BGMM.
(b) Transition matrices. Left: symmetric ground-truth transition matrix $p_{ij}^\mathrm{GT}$; Right: estimated transition matrix $\hat{p}_{ij}$. The asymmetric case is presented in Supplementary Section~\ref{SM-sec:appendix_asymmetric}.
}
\label{fig3}
\end{figure}

Using the simulation of the 3-state BM model, we numerically examine the performance of our model inference and parameter estimation algorithms.  We have simulated fifty trajectories of the 3-state BM model with $T=200$ and $\mathbf{D} = \{0.02, 0.08, 0.32\}~\mathrm{\mu m^2/s}$. 
First, we obtain a set of $\{\mathrm{M}_N, \mathbf{D} \}$ from the 50 trajectories using our diffusion-state analysis method. Next, we cluster the logarithm of the estimated diffusion coefficients by means of the Bayesian Gaussian Mixture Model (BGMM)~\cite{scikit-learn}. 
It is noted that a four-Gaussian component BGMM is employed in this analysis despite the simulated process being a 3-state BM model. We have confirmed that the redundant Gaussian component is eventually eliminated in the course of the BGMM analysis, and it clusters the simulation data with three Gaussian components.

Figure~\ref{fig3}(a) summarizes the results where the diffusion model has been inferred via the Bayesian selection model (see Fig.~\ref{SM-fig_a3} for the result from the AIC). Here, the right panel presents the clustering profile of the estimated diffusion coefficients via MLE and MAP, where the solid lines represent the ground-truth diffusion coefficients while the dashed lines indicate the inferred diffusion coefficients through BGMM.

Although the ground truth is the 3-state BM, inference often favors a 2-state model because short-lived states are difficult to resolve in finite trajectories. With a mean dwell time of $\sim$20 steps, brief visits in a $T=200$ trajectory provide too few samples for reliable estimation, so model selection criteria tend to merge them into a 2-state representation. Increasing the trajectory length to $T=500$ alleviates this issue, as more transitions and longer dwell segments allow the 3-state BM to emerge as the dominant outcome (Fig.~\ref{SM-fig_a5}; Supplementary Section~\ref{SM-sec:appendix_t500}).

Importantly, even when model selection favors the 2-state BM, the clustering of the estimated diffusion coefficients across all trajectories successfully predicts three states (Right panels in Fig.~\ref{fig3}(a) \& Fig.~\ref{SM-fig_a3}). Regardless of the Bayesian or AIC model selection methods, the clustered diffusivity values (dashed lines) via MLE or MAP agree well with the ground truth values (solid lines).


In our approach, we define the estimator of the diffusion coefficients as the average of the four distinct outputs from the two model inference methods and the subsequent two parameter estimation algorithms. This can be expressed as: 
\begin{eqnarray} \label{eq:D_estimate}
\begin{aligned}
\hat{\mathbf{D}} &= 
\frac{1}{4} \left ( \mathbf{D}^{\mathrm{Bayes}}_\mathrm{MLE}+\mathbf{D}^{\mathrm{Bayes}}_\mathrm{MAP}+\mathbf{D}^{\mathrm{AIC}}_\mathrm{MLE}+\mathbf{D}^{\mathrm{AIC}}_\mathrm{MAP}\right )\\
&= \{ \hat{D}^{(1)},~\hat{D}^{(2)},~\hat{D}^{(3)} \},
\end{aligned}
\end{eqnarray}
where $\mathbf{D}$ represents the mode values of the Gaussian components (dashed lines in Fig.~\ref{fig3}) in the clustering of diffusion coefficients via BGMM.

\subsection{\label{sec:transition_matrix}The transition matrix and segment identification}

After identifying the diffusion states, the next step is to construct the transition probability matrix among the diffusion states extracted from the trajectory data. 
The estimation of the transition matrix needs extra care compared to the task of identifying the diffusion states. The transition matrix cannot be obtained in the average sense with the individual transition matrices from single trajectories. The individual ones may have different sizes of $N \times N$, where $N$ denotes the number of diffusion states in a single trajectory. Furthermore, the statistical accuracy of the transition matrices obtained from an individual trajectory can be poor because the number of inter-state transition events necessary for estimating off-diagonal transition probabilities $p_{ij}$ $(i \neq j)$ is usually insufficient with a single trajectory. 

Our approach for estimating the transition matrix from a set of SPT data is the following: we start from an idea that a diffusing particle under investigation moves with distinct diffusion states $\hat{\mathbf{D}}=\{ D^{(1)}, \dots, D^{(N)} \}$, given by \eqref{eq:D_estimate}, and the observed trajectories are stochastic realizations of this process. We then conceptualize a long observation of this process by sequentially concatenating all the trajectories at hand. 
From this, we obtain the optimal values of $p_{ij}$ via MLE (Supplementary Section~\ref{SM-sec:appendixBW}). This estimation is denoted as
\begin{equation}
\hat {\mathbf{P}}_\mathrm{MLE}=
\begin{pmatrix}
\hat{p}_{11} & \hat{p}_{12} & \hat{p}_{13} \\
\hat{p}_{21} & \hat{p}_{22} & \hat{p}_{23} \\
\hat{p}_{31} & \hat{p}_{32} & \hat{p}_{33}
\end{pmatrix}.
\end{equation}
For estimating the components of $\mathbf{P}$, we solely rely on MLE, omitting the alternative method using MAP due to its computationally demanding nature.

We test the performance of our algorithm on a dataset with a symmetric transition matrix ($p_{ij}=p_{ji}$), as illustrated in Fig.~\ref{fig3}(b). The left panel presents the ground-truth transition probabilities used in the simulation, while the right panel shows the prediction $\hat{\mathbf{P}}_\mathrm{MLE}$. The relative error, $\frac{1}{N^2}\sum_{i,j}|p_{ij}^\mathrm{GT}-\hat{p}_{ij}|/p_{ij}^\mathrm{GT}$, is estimated to be $\approx 0.14$, demonstrating the excellent performance of our algorithm. The asymmetric transition matrix is also successfully inferred (Supplementary Section~\ref{SM-sec:appendix_asymmetric}).


Once the model parameters $\hat{\boldsymbol{\theta}}= \{ \hat{\mathbf{D}},  \hat{\mathbf{P}}_{\mathrm{MLE}}\}$ are determined, we identify the diffusion state of a trajectory data $x(t)$ as a function of $t$ by calculating the optimal diffusion state at every time step $t$ according to $\mathrm{argmax}_{D_t} P(D_t | \Delta x_{1:T}, \hat{\boldsymbol{\theta}})$.
Here the conditional probability $P(D_t | \Delta x_{1:T}, \hat{\boldsymbol{\theta}})$ is computed using the expectation step of the Baum-Welch algorithm [\eqref{SM-eq:FBalgorithm}].  

\section{\label{sec:MSHanalysis}The MSH sliding clamp exhibits three-state diffusion dynamics}

We now apply our single-trajectory Bayesian modeling tool to experimental SPT data of the MSH sliding clamp. 
Our analysis involves the identification of discrete diffusion states, the estimation of pertinent model parameters, segment identification, and a comprehensive characterization of the diffusion dynamics associated with each discerned diffusion state.

\subsection{MSH proteins have three distinct diffusion states }\label{sec:4a}

\begin{figure*}[!ht]
\centering
\includegraphics[width=0.85\textwidth]{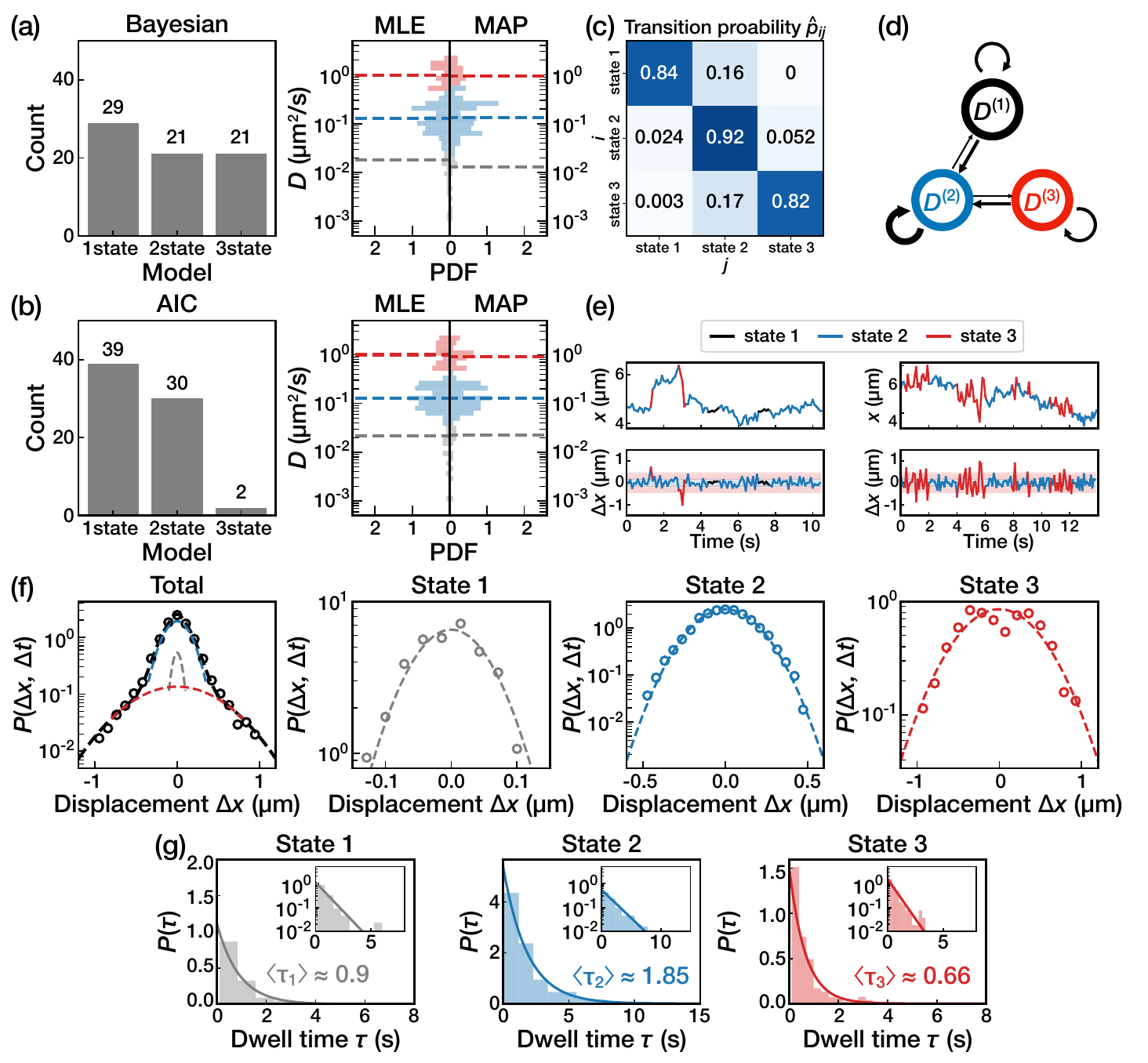} 
\caption{\scriptsize\textbf{Temporal heterogeneous diffusion of the MSH sliding clamp revealed by our diffusion-state analysis method.} (a) \& (b): The histogram of the inferred diffusion models $\mathrm{M}_N$ (Left) and the clustering of the estimated diffusion coefficients (Right). In the Right panels, the diffusion coefficients are inferred using MLE and MAP, and the dashed lines represent the modes of the Gaussian components found by BGMM. 
(a) Results from the Bayesian selection model. (b) Results from the AIC.
(c) Transition matrix among three dynamic states found from the SPT data. (d) Transition model based on the estimated transition matrix in (c). The flow diagram with the arrows having different thickness illustrates the dominant transition pathway.  
(e) Two representative MSH trajectories with the indication of the three dynamic states identified by our analyses. The upper panels display the time series of the position of an MSH while the lower panels show the respective displacement time series. The Baum-Welch algorithm is employed to identify the dynamic state along a trajectory. Black: state 1 (immobile state) having a diffusion coefficient $\hat{D}^{(1)} \approx 1.86 \times 10^{-2}~\mathrm{\mu m^2/s}$. Blue: state 2 having a diffusion coefficient $\hat{D}^{(2)} \approx 1.30 \times 10^{-2}~\mathrm{\mu m^2/s}$. Red: state 3 having a diffusion coefficient $\hat{D}^{(3)} \approx 9.64 \times 10^{-1}~\mathrm{\mu m^2/s}$. In the lower panels, the blue and red shades indicate the standard deviations for state 2 ($\sqrt{2 \hat{D}^{(2)} t_0} \approx 0.16~\mathrm{\mu m}$) and state 3 ($\sqrt{2 \hat{D}^{(3)} t_0} \approx 0.44~\mathrm{\mu m}$), respectively.
(f) Displacement distributions of the MSH trajectories. The leftmost panel shows the displacement distribution at $\Delta t=t_0$ for the original trajectories. In this plot, the gray, blue, and red dashed lines depict the Gaussian components of state 1, state 2, and state 3, respectively. The black dashed line represents the sum of these three Gaussian functions. The three panels on the right show the normalized displacement distributions for the three diffusion states (symbols), together with the corresponding Gaussian curves whose standard deviations are given by $\sigma^{(i)} = \sqrt{2 \hat{D}^{(i)} t_0}$ (dashed lines).
(g) Dwell time distributions of states 1 (gray), 2 (blue), and 3 (red) with the best exponential fit (solid lines). The average dwell time $\langle \tau_i\rangle$ is indicated in the plot as the characteristic time of the exponential distribution. The inset shows the log-linear plot of the same data.
}
\label{fig4}
\end{figure*}

\begin{table*}
\centering
\caption{Parameters of Gaussian components inferred from BGMM applied to the estimated parameters via MLE and MAP, shown in the right panels of Figs.~\ref{fig4}(a) \& (b).}
\label{table:table1}
{\footnotesize
\begin{tabular}{cc|cc|cc}
\noalign{\smallskip}\noalign{\smallskip}\hline\hline
\multirow{2}{*}{} & & \multicolumn{2}{c|}{Bayesian} & \multicolumn{2}{c}{AIC} \\
\cline{3-6}
     & & MLE  & MAP & MLE & MAP \\
\hline
\multirow{3}{*}{weight}
&$w_1$&
$0.1680$ & $0.1115$ &
$0.1472$ & $0.1460$ \\
&$w_2$&
$0.5824$ & $0.6728$ &
$0.5995$ & $0.6076$ \\
&$w_3$&
$0.2417$ & $0.2073$ &
$0.2433$ & $0.2364$ \\
\hline
\multirow{3}{*}{\shortstack[1]{mode\\ $~(\mathrm{\mu m^2/s})$}}
&$\mu_1$&
$1.80\times10^{-2}$ & $1.28\times10^{-2}$ &
$2.14\times10^{-2}$ & $2.22\times10^{-2}$ \\
&$\mu_2$&
$1.29\times10^{-1}$ & $1.35\times10^{-1}$ &
$1.28\times10^{-1}$ & $1.29\times10^{-1}$ \\
& $\mu_3$&
$9.92\times10^{-1}$ & $9.51\times10^{-1}$ &
$9.96\times10^{-1}$ & $9.17\times10^{-1}$ \\
\hline
\multirow{3}{*}{variance}
& $\sigma_1^2$&
$0.3701$ & $0.4534$ &
$0.3774$ & $0.4017$ \\
& $\sigma_2^2$&
$0.1009$ & $0.1396$ &
$0.0694$ & $0.0768$ \\
& $\sigma_3^2$&\
$0.0649$ & $0.0584$ &
$0.0623$ & $0.0597$ \\
\hline
\hline
\end{tabular}
}
\end{table*}

In Figs.~\ref{fig4}(a) and (b), we present the results of the model selection and diffusion coefficient estimation. The left panels display the histogram of the optimal multi-state BM models for individual SPT datasets. On the right, we depict the clustering of estimated diffusion coefficients through two estimators, MLE and MAP, using a four-Gaussian component BGMM. A significant finding is that the diffusion coefficients exhibit clear clustering into three major Gaussian components, with a negligible weight ($w_4 < 0.03$) assigned to the fourth component (not shown). The dashed lines in the clustering plot (right panels) represent the three Gaussian components. Refer to Table~\ref{table:table1} for detailed information on these three diffusion states.
  
Figure~\ref{fig4}(a) shows the MSH clamp's diffusion state revealed by our Bayesian selection method. As previously mentioned, the MSH clamp exhibits a maximum of three distinct states within our observation time window. However, a considerable portion of trajectories captures single or two-state diffusion. The clustering of diffusion coefficients, estimated by both MLE and MAP, is well-captured by three Gaussian components. The diffusion coefficients of these three states are determined by the mode values of each Gaussian component. By averaging the mode values from MLE and MAP (refer to Table~\ref{table:table1}), we identify that the MSH clamp possesses diffusion coefficients approximately equal to $\mu_3 \sim 9.6 \times 10^{-1}\mathrm{\mu m^2/s}$, $\mu_2 \sim 1.3 \times 10^{-1}\mathrm{\mu m^2/s}$, and $\mu_1 \sim 1.9 \times 10^{-2}~\mathrm{\mu m^2/s}$.

We validate these findings by employing the AIC-based model selection, as illustrated in Fig.~\ref{fig4}(b). Notably, the AIC method tends to detect the two-state BM more than the Bayesian model selection. Despite this disparity, the clustering of diffusion coefficients yields highly consistent results. Three distinct diffusion states are evident, and their respective diffusion coefficients are in good agreement with those obtained through the Bayesian method [Table~\ref{table:table1}]. 

To sum up, we conclude that three distinct diffusion states exist in the MSH clamp's diffusion along DNA. Designating these states as state 1 (gray cluster), state 2 (blue), and state 3 (red), the point estimators of the corresponding diffusion coefficients result in
\begin{eqnarray}\label{eq:D_values}
\begin{aligned}
&\hat{\mathbf{D}}=(\hat{D}^{(1)}, \hat{D}^{(2)}, \hat{D}^{(3)})\\
&\approx (1.86 \times 10^{-2},~1.30 \times 10^{-1},~9.64 \times 10^{-1} ) ~\mathrm{\mu m^2/ s}.
\end{aligned}
\end{eqnarray}
Among the three diffusion states, the diffusion coefficient of state 2 is of similar order to those reported for the sliding clamps of MutS homologs~\cite{britton2022exploiting, london2021linker, cho2012}.
State 1 is hypothesized to represent aa slow phase, potentially indicative of MSH binding to either a specific or nonspecific DNA sequence. Its diffusion coefficient, $\hat{D}^{(1)}$, is lower than those observed in any known diffusive modes of MSH.
It seems that state 3 is a newly identified diffusion state revealed by our SPT experiment and single-trajectory Bayesian tool. The corresponding diffusion coefficient, $\hat{D}^{(3)}$, has not been reported in the literature previously.

\subsection{Transition probability and segment identification}\label{sec:4b}

After the identification of the three diffusion states along with their respective diffusion coefficients, we estimate the transition probabilities, $\hat{p}_{ij}$, among these states via our methodology:
\begin{eqnarray}\label{eq:P_est}
\hat{\mathbf{P}}_{\mathrm{MLE}} \approx
\begin{pmatrix}
0.8372 & 0.1628 & 0.0000 \\
0.0244 & 0.9232 & 0.0525 \\
0.0026 & 0.1724 & 0.8250
\end{pmatrix}.
\end{eqnarray}
Subsequently, based on this transition matrix, we calculate the stationary distribution for each state 
\begin{eqnarray}\label{eq:stationary_ratio}
\hat{\boldsymbol{\pi}}_{\mathrm{MLE}} \approx (0.1062,~0.6876,~0.2062)^\intercal.
\end{eqnarray}
It is noteworthy that transitions between states 1 and 3 are notably rare. Specifically, the observed probabilities suggest that $p_{13} \ll p_{12}$ and $p_{31} \ll p_{32}$, leading us to propose that state 2 potentially acts as a mediator between states 1 and 3. Figure~\ref{fig4}(d) provides a visual representation of the transition pathway among these three states.

Equipped with the estimated parameters $\hat{\boldsymbol{\theta}}= { \hat{\mathbf{D}}, \hat{\mathbf{P}}_{\mathrm{MLE}}}$, we apply the segment identification method to the original SPT data. Figure~\ref{fig4}(e) showcases two sample MSH trajectories, illustrating the identified dynamic states of an MSH as a function of time. Indeed, an MSH sliding clamp temporally switches its diffusion states while moving along DNA. Typical dwell times of each state are on the order of seconds, with states 2 and 3 appearing more prominently than state 1 in the trajectories, consistent with the features indicated by the transition matrix and stationary distribution [\eqref{eq:P_est} \& \eqref{eq:stationary_ratio}].

\subsection{Characterization of distinct diffusion states of MSH sliding clamp}\label{sec:4c}

Using the state-labeled trajectories, we conduct a statistical analysis of diffusion characteristics. Firstly, we calculate the van-Hove self-correlation function separately for the three states. In Fig.~\ref{fig1}(e), we observe that the displacement distribution from the collection of the three contributions is non-Gaussian. Since our diffusion-state analysis concludes that the MSH diffusion consists of three distinct Brownian states, we anticipate that the displacement distribution at $\Delta t = t_0$ is constructed by the superposition of three Gaussian propagators (i.e. components), where each Gaussian component corresponds to the respective diffusion state with $\hat{D}^{(i)}$ found in \eqref{eq:D_values}. By counting the number of unit-time displacements of each diffusion state, we obtain their relative populations $c_1$, $c_2$, and $c_3$ with $\sum_{i=1}^{3} c_i = 1$. Our estimation finds $c_1 \approx 0.0760$, $c_2 \approx 0.7555$, and $c_3 \approx 0.1685$, which are similar to the stationary distribution in \eqref{eq:stationary_ratio}. The van-Hove self-correlation function for the original data is then described by
\begin{eqnarray} \label{eq:threeGaussian}
P(\Delta x, t_0) = \sum_{i=1}^{3} c_i \frac{\exp \left (-\frac{\Delta x^2}{4\hat{D}^{(i)} t_0} \right )}{\sqrt{4\pi \hat{D}^{(i)} t_0}}.
\end{eqnarray}
As demonstrated in Fig.~\ref{fig4}(f), the superposition (dashed line) of the three Gaussian propagators via \eqref{eq:threeGaussian} excellently explains the experimental data (black dots).

Additionally, we confirm that the estimated van-Hove self-correlation function of each state (right three panels in Fig.~\ref{fig4}(f)) is Gaussian, consistent with our multi-state BM model. In each panel, the data (symbol) agree closely with the Gaussian curves (dashed lines), with standard deviations given by $\sigma^{(i)} = \sqrt{2 \hat{D}^{(i)} t_0}$. The Gaussian nature of each diffusion state persists at longer lag times, see Fig.~\ref{SM-fig_a6} (Supplementary Section~\ref{SM-sec:displacement_distribution}).

In Fig.~\ref{fig4}(g), we obtain the dwell time distribution for the three dynamic states using the state-identified trajectories. Assuming Markovian transitions, the dwell time is expected to follow an exponential distribution. The dwell time distribution for state $i$, with mean dwell time $\langle \tau_{i} \rangle$, is given by
\begin{eqnarray}\label{eq:geometric}
P(\tau) = \frac{1}{\langle \tau_i \rangle} \exp \left ( -\frac{1}{\langle \tau_i \rangle} \tau \right ).
\end{eqnarray}
Consistent with expectations, the exponential distribution \eqref{eq:geometric} explains reasonably the experimental data. The mean dwell times for the three states are estimated to be $\langle \tau_1 \rangle \approx 0.9~\mathrm{s}$, $\langle \tau_2 \rangle \approx 1.85~\mathrm{s}$, and $\langle \tau_3 \rangle \approx 0.66~\mathrm{s}$. Notably, the fastest state (state 3) exhibits the shortest mean dwell time among the three, while the mediator state (state 2) has the longest.

\section{Discussion and conclusions}

\begin{figure*}[ht]
\centering
\includegraphics[width=0.6\textwidth]{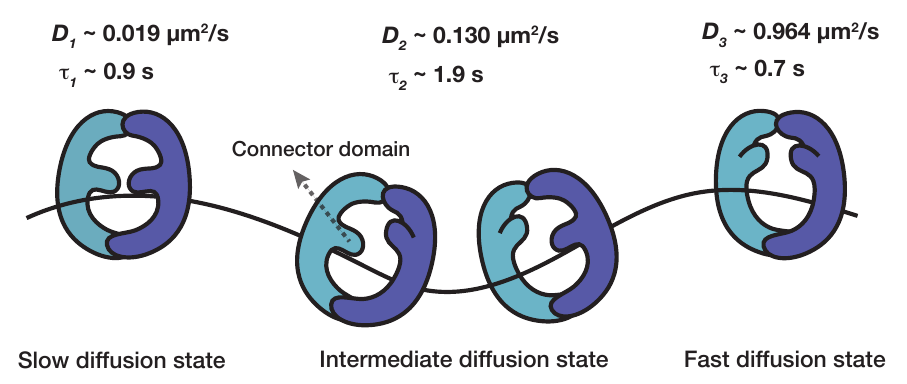} 
\caption{\textbf{A toy model illustrating distinct diffusion states of the MSH sliding clamp.} State 1 (Left): Strong interactions with DNA significantly slow down the MSH diffusion. State 2 (Middle): One of the DNA-binding interfaces in the connector domains contact DNA, resulting in two distinct sub-states. State 3 (Right): The connector domains are in a contact-free state with DNA, resulting in the fastest 1D diffusion of MSH. 
}
\label{fig5}
\end{figure*}

Through comprehensive single-trajectory Bayesian analyses, we revealed the temporal heterogeneity in the diffusion dynamics of the human MSH sliding clamp. To rigorously quantify the observed stochasticity and multi-state behavior, we applied a hidden Markov model grounded in Bayesian inference. Our new approach enabled us to determine the number of distinct diffusion states, their diffusion coefficients, and the transition pathways among them. The validity and robustness of our diffusion-state analysis were further confirmed using simulated datasets. 

A primary finding of our study is that the MSH sliding clamp possesses three distinct diffusion states, characterized by the diffusion coefficients of about $\hat{D}^{(1)}=1.86 \times 10^{-2}$, $\hat{D}^{(2)}=1.30 \times 10^{-1}$ and $\hat{D}^{(3)}=9.64 \times 10^{-1}~(\mathrm{\mu m^2/s})$, respectively. Further examination of these diffusion states revealed that each state follows Gaussian diffusion, with a dwell time of $\sim 1~\mathrm{s}$ and frequent transitions between the diffusion states. The dominant portion of state transitions occurred between states 2 (intermediate diffusion) and 3 (fast diffusion). Conversely, transitions entering into state 1 (slow diffusion) were found to be scarce. We emphasize that the non-Gaussian van-Hove self-correlation observed in the entire displacement dataset originates from this temporally heterogeneous diffusion dynamics. The non-Gaussian nature was accounted for by the superposition of three Gaussian van-Hove self-correlations from states 1--3. Our model selection and estimation of diffusion coefficients indicated that the majority of trajectories showcase homogeneous diffusion behavior, with the estimated diffusion coefficients aligning with that of state 2 in an ATP-rich solution. 

To discuss the mechanisms underlying the three diffusion states of the MSH sliding clamp, we assume that its structural basis can be inferred from the \emph{E. coli} MutS sliding clamp bound to MutL, given the high conservation of MutS from prokaryotes to eukaryotes. A crystal structure of the MutS-MutL (the $40~\mathrm{kDa}$ N-terminal LN40 domain) complex was resolved in the presence of the non-hydrolyzable ATP analog AMP-PNP (adenylyl-imidodiphosphate) and a G/T mismatch ~\cite{groothuizen2015muts}. This structure revealed that the connector domains, containing DNA and MutL-binding interfaces, move outward from the mismatch site. This conformational change induced by ATP binding enables the recruitment of MutL to DNA. Notably, to our knowledge, a crystal structure of the ATP-bound MutS sliding clamp in the absence of MutL has not been observed. Taken together, we suspect that the LN40 domain of MutL captures the fluctuating connector domains as they transition between DNA binding and unbinding states, thereby stabilizing the complex.  

Based on this structural insight, we hypothesize that distinct diffusion states of the MSH sliding clamp arise from conformational changes associated and dissociated with the DNA-binding status of the connector domains (Fig.~\ref{fig5}): (1) Fastest diffusion state (state 3): both connector domains are in an open conformation, minimizing contact between the MSH clamp and DNA. (2) Intermediate diffusion state (state 2): one of the connector domains engages with DNA. (3) Slow diffusion state (state 1): both DNA-binding domains contact DNA. This slow diffusion state frequently occurs when the sliding clamp is located far from the mismatch (Supplementary Section~\ref{SM-sec:mismatch}), in contrast to a previous report suggesting that Taq MutS sliding clamp can revisit DNA mismatches during diffusion~\cite{hao2020recurrent}. This speculation is consistent with our analyses, including the transition probabilities between diffusion states (Fig.~\ref{fig4}(c) and \ref{fig4}(d)) and the dwell times of each state (\eqref{eq:geometric}). Because state 2 comprises two sub-states, transitions between state 1 and 3 occur via state 2, which also exhibits a longer dwell time than the other states. 

Our single-trajectory Bayesian framework can be widely employed to unveil the diffusion dynamics of other DNA-binding proteins or intracellular macromolecules. For example, SPT-based experimental studies have reported temporal- and particle-to-particle heterogeneous diffusion dynamics in DNA-binding proteins~\cite{beckwitt2020single, cheon2019single, kamagata2018high, lin2014, kong2016single}. However, these investigations often lack rigorous quantitative analysis of the observed diffusion heterogeneity and stochasticity. Our methodologies provide robust and quantitative assessments of diffusion heterogeneity. We envision that our work can contribute to the comprehensive understanding of complex diffusion dynamics of DNA-binding proteins, facilitating deeper insights into their functional mechanisms.

\section*{Methods}
\subsection{Construction of mismatch-containing and labeled DNA for DNA skybridge}

The $\lambda$-phage DNA (48.5 kb; New England Biolabs) molecules with a single G/T mismatch were constructed by CRISPR/Cas9. To create the oligo exchange site, $\lambda$-phage DNA was treated with a CRISPR crRNA set (‘5-AAUUAAGGGUUACUAUAUGU-3’, ‘5-UGUUGCCGCCAAAUAAAUUG-3’) and tracrRNA (IDT), which were annealed at 80°C for 5 minutes and then slowly cooled to form the sgRNA. Next, Cas9 nickase (New England Biolabs) was added, and the reaction was incubated at room temperature for 20 minutes to form the RNP complex. The $\lambda$-phage DNA was then incubated at 37°C for 1 hour for nicking. Following the nicking reaction, CRISPR/Cas9 activity was stopped using RNase and Proteinase K (20 mg/mL; Thermo Scientific), and the remaining components were removed using an Amicon filter(MWCO $100~\mathrm{kDa}$).
The oligo exchange was then performed using a DNA oligo containing both the mismatch and a fluorescent label (‘5-phosphate/TATAGTAACCCT/iAlexa488N /AATTTTATTAGAATAACCGCAA-3’; Integrated DNA Technologies), which was annealed at 80°C for 5 minutes. Finally, T4 DNA Ligase (10U, Roche) was used for ligation at 18°C overnight. The biotin-attached oligos (’5-phosphate/AGGTCGCCGCCCTT-3’biotin) and (5’-phosphate/GGGCGGCGACCTTT-3’biotin) were mixed with the $\lambda$-phage DNA to form bi-biotin DNA substrates. The annealing process was performed for 5 minutes at 80°C and the cooling process was done right after. T4 DNA Ligase (10U, Roche) was used for ligation at 18°C overnight. For the final purification process, Float-A-Lyzer (Spectra-Por$^{\circledR}$ Float-A-Lyzer$^{\circledR}$ G2 blue, 1 mL, MWCO 100kDa) was used to filter unbound oligos.

\subsection{Single-molecule total internal reflection fluorescence microscopy with DNA Skybridge light sheet imaging}

The real-time imaging of protein diffusion was collected by single-molecule total internal reflection fluorescence (smTIRF) imaging with DNA skybridge~\cite{kim2019dna}. For the skybridge pattern quartz, photolithography processes were performed on (100) quartz to make a height difference between the surface and imaging field. The 1.6X magnifier in a prism-type TIRF microscope (Olympus IX-71, water-immersion 60X objective NA = 1.2), EMCCD (ImagEM C9100-13, Hamamatsu), and MetaMorph 7.6 (Molecular Devices) imaging software were used for real-time imaging. The diffusion of MSH2-Alexa647-MSH6 was imaged using a $532~\mathrm{nm}$ laser, and the Alexa488 at the mismatch site was excited with a $488~\mathrm{nm}$ laser at a $100~\mathrm{ms}$ frame rate.
The DNA skybridge functionalized with polyethylene glycol (PEG) and PEG-biotin was coated by incubating the surface for 16 minutes with blocking buffer ($20~\mathrm{mM}$ Tris-HCl, pH 7.5, $2~\mathrm{mM}$ EDTA, $50~\mathrm{mM}$ NaCl, and 0.0025~\% Tween 20 (v/v)) containing BSA ($0.2~\mathrm{mg/mL}$; NEB). Streptavidin ($0.05~\mathrm{mg/mL}$, Sigma-Aldrich) was then incubated for 10 minutes, followed by washing with blocking buffer. Biotin bi-tethered mismatch DNA ($48.5~\mathrm{kbp}$) at $100~\mathrm{pM}$ was flowed into the system using a syringe pump at a rate of $90~\mathrm{\mu L/min}$.
The labeled MSH2-Alexa647-MSH6 was used to observe single-particle diffusion on the DNA skybridge, with molecules incubated at $2~\mathrm{nM}$ concentrations for 2 minutes in a solution containing $30~\mathrm{mM}$ Tris-HCl (pH 7.5), $100~\mathrm{mM}$ potassium glutamate, $5~\mathrm{mM}$ MgCl2, $0.1~\mathrm{mM}$ DTT, $1~\mathrm{mM}$ ATP, $0.2~\mathrm{mg/mL}$ acetylated BSA, 0.0025~\% tween 20, $1~\mathrm{mM}$ PCA, $10~\mathrm{nM}$ PCD, and $2~\mathrm{mM}$ Trolox. After 2 min of incubation time, the diffusion was recorded with no extra solution exchange.

\subsection{Data availability}
XXX

\subsection{Code availability}
XXX (if you have)

\section*{Author contributions}
S.P. contributed to the development of theoretical modeling, analysis of experimental trajectories, simulations, and the writing of the manuscript. I.Y. performed the experiment and produced the data. J.L., S.K., and J. M.-L. contributed to the experimental setup and the protein preparation. R.F. provided the proteins and supervised the experiment. J.B. conceived and supervised the experiment and contributed to the writing of the manuscript. J.-H.J conceived the research, supervised the theoretical modeling \& analyses, and wrote the manuscript. 

\begin{acknowledgments}
This work was supported by the National Research Foundation
(NRF) of Korea, Grant No.~2021R1A6A1A10042944 \& No. RS-2023-00218927.
\end{acknowledgments}


\bibliography{references}

\end{document}